\documentclass[12pt,preprint]{aastex}

\def\aa{{A\&A}}

\def\aj{{AJ}}

\def\apj{{ApJ}}
\def\apjs{{ApJS}}

\def\gax{{$\mathrel{\hbox{\rlap{\hbox{\lower4pt\hbox{$\sim$}}}\hbox{$>$}}}$}}
\def\hal{H$\alpha$}

\def\lax{{$\mathrel{\hbox{\rlap{\hbox{\lower4pt\hbox{$\sim$}}}\hbox{$<$}}}$}}

\def\mnras{{MNRAS}}

\def\percm2{cm$^{-2}$}

\shorttitle{Bluetails }
\shortauthors{Neff et al.}

\begin{document}


\title{UV Emission from Stellar Populations within Tidal Tails:  \\
    Catching the Youngest Galaxies in Formation?}


\author{S. G.  Neff\altaffilmark{1}, D. A.  Thilker\altaffilmark{2},
M. Seibert\altaffilmark{3}, A. Gil de Paz\altaffilmark{4},
L.  Bianchi\altaffilmark{2},  D. Schiminovich\altaffilmark{3,5}
C. D. Martin\altaffilmark{3},   B. F.  Madore\altaffilmark{4},
R. M. Rich\altaffilmark{6}, T. A. Barlow\altaffilmark{3}, Y-I.
Byun\altaffilmark{7}, J. Donas\altaffilmark{8}, K. Forster\altaffilmark{3}, P.
G. Friedman\altaffilmark{3}, T. M. Heckman\altaffilmark{9}, P. N.
Jelinsky\altaffilmark{10}, Y-W. Lee\altaffilmark{7}
, R. F. Malina\altaffilmark{8}, B. Milliard\altaffilmark{8}, P.
Morrissey\altaffilmark{3}, O. H. W. Siegmund\altaffilmark{10}, T.
Small\altaffilmark{3}, A. S. Szalay\altaffilmark{9}, B. Y.
Welsh\altaffilmark{10}, and T. K. Wyder\altaffilmark{3}}

\altaffiltext{1}{Laboratory for Astronomy and Solar Physics,
NASA Goddard Space Flight Center, Greenbelt, MD  20771,susan.g.neff@nasa.gov}
\altaffiltext{2}{Center for Astrophysical Sciences, The Johns Hopkins
University, 3400 N. Charles St., Baltimore, MD 21218}
\altaffiltext{3}{California Institute of Technology, MC 405-47, 1200 East
California Boulevard, Pasadena, CA 91125}
\altaffiltext{4}{Observatories of the Carnegie Institution of Washington,
813 Santa Barbara St., Pasadena, CA 91101}
\altaffiltext{5}{Columbia University, New York, NY 10027}
\altaffiltext{6}{Department of Physics and Astronomy, University of California
at Los Angeles, Los Angeles, CA 90095}
\altaffiltext{7}{Center for Space Astrophysics, Yonsei University, Seoul
120-749, Korea}
\altaffiltext{8}{Laboratoire d'Astrophysique de Marseille, BP 8, Traverse
du Siphon, 13376 Marseille Cedex 12, France}
\altaffiltext{9}{Department of Physics and Astronomy, The Johns Hopkins
University, Homewood Campus, Baltimore, MD 21218}
\altaffiltext{10}{Space Sciences Laboratory, University of California at
Berkeley, 601 Campbell Hall, Berkeley, CA 94720}

\begin{abstract}

New GALEX observations have detected significant FUV (1530\AA)
and NUV (2310\AA) emission from stellar substructures
within the tidal tails of four ongoing galaxy
mergers.  The UV-bright regions are optically faint and
are coincident with HI density enhancements.  FUV emission
is detected at any location where the HI surface density
exceeds $\sim$ 2 M$_{\odot}$ pc$^{-2}$, and is often detected
in the absence of visible wavelength emission.
UV luminosities of the brighter regions of the tidal tails
imply masses of 10$^6$M$_{\odot}$ up to $\sim$ 10$^9$M$_{\odot}$ 
in young stars in the
tails, and HI luminosities imply similar HI masses.
UV-optical colors of the tidal tails indicate stellar populations
as young as a few Myr, and in all cases ages $<$ 400Myr.
Most of the young stars in the tails formed in single bursts
rather than resulting from continuous star formation, and they
formed {\it in situ} as the tails evolved.  Star
formation appears to be older
near the parent galaxies and younger at increasing distances from the
parent galaxy.  This could be because the star formation
occurs progressively along the tails, or because the
star formation has been inhibited near the galaxy/tail interface.
 The youngest stellar
concentrations, usually near the ends of long tidal
tails, have masses comparable to confirmed tidal dwarf galaxies
and may be newly forming galaxies undergoing their first burst
of star formation.

\end{abstract}




\section{Introduction}\label{sec:intro}

   Galaxy evolution is strongly influenced by interactions and
mergers.  Dramatic structural changes often result from such events.
Interactions and mergers occur more frequently at higher
redshifts, and were even more important in the
early universe than today.

    Tidal tails may be ejected $> 100$ kpc during strong
interactions, with more than half of the neutral gas in each
galaxy ending up in the tail \citep{bra01}.  The tails often are
blue \citep{sch90}; occasionally \hal\ emission and young stars
(few 100 Myr) are detected, indicative of active star-formation
\citep{hib96, kni03}. Models predict that gas condensations in
tidal tails will self-gravitate into new ``tidal dwarf galaxies"
(TDGs) \citep{bar92, mih94, mih96}; such predictions are confirmed
by observation \citep{mir92, duc98, bra01}.  Recent work on tidal
tails \citep{sav04, kni03, deg03} has established that young star
clusters and associations (10-500 Myr) can form in tidal tails.

   In this Letter, we present analysis of strong UV emission
associated with tidal tails in four major galaxy interactions.
Hibbard et al. (this volume) also report the GALEX detection of
{\it in situ} star formation within the tidal tails of the Antennae
(NGC~4038/39).

\section{Targets, Observations, and Data Analysis} \label{sec:observations}

\subsection{Targets} \label{sec:Targets}

We present GALEX UV imagery (Figure 1, a-d), plus archival 
visible-band and HI
observations (Figure 1, e-h), for NGC~7769/71, 
NGC~5713/19, Arp~295, and the NGC~520
(Arp~157) system.  They represent a range of merger stages, from
very early (NGC 7769/71 and NGC 5713/19), through early (Arp 295),
to nearly merged (NGC~520).

The NGC~7769/71 pair ($D$=56Mpc) dominate
a loose galaxy group embedded in a common HI envelope \citep{nor97}.
NGC~7771 is part of a triplet of galaxies including NGC~7770
and NGC 7771a.   A ring of higher column density encircles the
NGC~7771 triplet , and lower column density H~I
extends in a tail ($\ga 100$ kpc to the south).  \citet{nor97} suggest
that the galaxies have passed perigalacticon once in a
prograde (NGC~7771) - retrograde(NGC~7769) interaction, in which
a significant tidal tail was ejected from NGC7771, but very
little matter was removed from NGC~7769.

NGC~5713/19 ($D$=25Mpc) are a pair of interacting
galaxies that form part
of the larger galaxy group CfA139 \citep{hay91}.
Both galaxies have
extensive HI envelopes and tidal tails \citep{lan01}. \citet{hib01}
classify this system as a prograde - prograde interaction, where
two tidal tails have been ejected from gas-rich galaxies.
HI column densities are quite high in the tails, $4-8 \times 10^{20}$
cm$^{-2}$.   We focus on NGC~5719's tidal tail, which
extends $\ga$ 40 kpc.

The galaxies in the Arp~295 pair (D$=$69Mpc)
are the largest spirals in a loose group of galaxies.   They
appear connected by a dramatic optical and H~I bridge.
A luminous tail extends $\ga$ 50 kpc to the southwest of Arp~295a,
and a plume extends $\sim$ 35 kpc to east of Arp 295b
\citep{hib96}.  The system apparently resulted from a prograde
(Arp~295a) - retrograde (Arp~295b) encounter, when an extensive
tail was ejected from Arp295a (it wraps around to {\it appear} as the
bridge between the galaxies before extending off to the SW) and the
gas-rich plume east of Arp~295b was removed but remained near the
parent galaxy.

The NGC~520 system (D$=$23Mpc) is  more evolved, with the
progenitor disks
almost merged; K-band peaks indicate two nuclei separated by
only $\sim$5~kpc \citep{sta91,kot01}.
An optical and HI tidal tail extends $\sim$25kpc to the south
from the nuclei, then abruptly bends into a faint red plume and
H~I arm extending east and then north.  These structures are embedded in
a disk-like HI distribution \citep{hib00}, which extends
northward to the dwarf galaxy UGC~957 ($>$ 50kpc separation) and
then continues around to the west and south.
The NGC~520 system probably resulted from a prograde - retrograde
(or polar) interaction between two large disks \citep{sta91}.
\citet{hib00} suggest that UGC~957 may be a tidal dwarf galaxy
formed in the interaction.

\subsection{UV Observations} \label{sec:uvobsv}

   GALEX observed each system for a single orbit in both
FUV (1530\AA) and NUV (2310\AA) bands.  Routine data processing
and calibration was as described in \citet{mar04}.  Figure 1
shows the GALEX images compared to visible and $\lambda$21cm H~I imagery.
UV emission is seen to be associated with all of the interacting
galaxies and their tidal tails, extending large distances (10-100kpc)
beyond the optical extent.  The images were adaptively smoothed, and
morphologically distinct regions were
selected for photometry (FUV images were used to define apertures
at all wavelengths (Figure 1, top).  Aperture photometry was performed
using custom IDL routines; sky was determined as the mean of 8$-$12
boxes around the target.
Background-subtracted AB magnitudes, photometric errors, and
FUV$-$NUV colors were then determined (Table 1) from the original
unsmoothed data.  

The NGC~7769/71 GALEX image (1663 sec, Fig 1a) shows strong UV
features associated with the both principal galaxies, their satellites,
and with a peak in the SW HI tail of NGC~7771 at NGC~7771b.
A faint UV halo extends W of NGC 7771 towards NGC 7770 and NGC
7771a, in agreement with the extent of the HI.  NGC~7771b has the
most extreme FUV$-$NUV value in our galaxy sample.

The disks of NGC~5719 and NGC~5713 (not shown) are easily detected
in the short GALEX observation (844 sec, Fig 1b).  UV emission
extends well beyond NGC~5719's optical disk, continuing nearly the
entire length of NGC~5719's HI tidal tail ($\ga$ 40 kpc from the
nucleus).  The tail FUV emission consists of very blue
clumps at the locations of HI peaks, as well as a more diffuse
component.  Very faint bridge UV emission is also detected
coincident with the H~I NW of NGC~5719, extending towards NGC~5713.

UV emission is detected from several components of the Arp~295 system
(1234 sec, Figure 1c).
Arp~295b is a bright, compact source, while Arp~295a appears clumpy.
UV emission is detected from Arp~295a's tidal tail and from the
bridge,  however the tidal tail does not extend as far to the SW
in the UV as it does at visible wavelengths,
and the UV bridge fades in the middle.  The
plume east of Arp~295b is conspicuous in the UV, with
Clump~1 (embedded in the plume) the bluest UV location in the Arp~295 system.

The UV observations of NGC~520 (1730 sec, Figure 1d) show
that the central region appears similar at UV and visible
wavelengths \citep{hib00}), with FUV emission detected from
the remnant galaxy bodies.  The inner regions of two
clumpy tidal tails are clearly visible in the UV images, extending
SE and NW.  UGC~957, the likely TDG, shows up very clearly
in the FUV image, and has the bluest UV colors in
the system.  The red plume extending E and
N to UGC~957 is not detected in the UV.

\subsection{HI Observations} \label{sec:HIobsv}

   HI maps of each interacting system were obtained using the
NRAO VLA\footnote{The National Radio Astronomy Observatory is a facility
of the National Science Foundation operated under cooperative
agreement by Associated Universities, Inc.}.
Details of the array configuration, correlator setup, imaging
parameters, and subsequent deconvolution are in the original
papers (NGC~7769/71: \citet{nor97}, NGC~5713/19: \citet{lan01},
Arp 295 and NGC~520: \citet{hib96})\footnote{Thanks to T. Nordgren,
   G. Langston, and J. Hibbard for sharing their data.}.
Generally, the synthesized beam measured
$\sim 30\arcsec$ with integrations sufficient to accurately map HI
column densities $>$ few $\times 10^{19}$cm$^{-2}$.

   The HI distribution is seen to be closely matched to the FUV
emission, (Figure~1). The H~I and the FUV emission generally peak
at the same locations in the tidal tails (within the accuracy of
the H~I resolution).  The HI surface density generally exceeds
$\sim$ 2 M$_{\odot} pc^{-2}$ where FUV emission is detected (Table
1).  In locations where the H~I is below this level, FUV emission
is usually not detected.  This suggests that {\it current} star
formation is directly related to the cold gas density, and that a
threshold density must be reached before star formation can
begins.  Assuming that CO is coincident with H~I suggests that the
observed (H~I) density of $\sim$ 2 M$_{\odot}$ pc$^{-2}$ is a lower
limit to the star-formation threshold density in tidal tails.  {\it Higher
resolution H~I and spatially resolved CO images would be very
helpful in exploring this possibility.}

\subsection{Visible-Band Observations} \label{sec:visobsv}

   Photometry at visible wavelengths was performed for the
selected UV-bright regions (sky was determined from multiple 15''
$\times$ 15'' squares, fluxes of individual pixels were summed
using IRAF).  For NGC~7771 we used optical images in the SDSS
$gri$ bands from the INT\footnote{The Isaac Newton Telescope is
operated on the island of La Palma by the Isaac Newton Group in
the Spanish Observatorio del Roque de los Muchachos of the
Instituto de Astrof\'{\i}sica de Canarias.} Wide Angle Survey
\citep{McM99}, and for NGC~5719 we used Sloan Digital Sky Survey
(SDSS) $ugriz$ images.  Optical images for Arp~295 ($BVR$) and
NGC~520 ($V$) were obtained by Hibbard \& van Gorkom (1996)
  \footnote{Thanks to J. Hibbard for making the calibrated data
  available.}.
After removal of field stars we computed apparent magnitudes with
associated errors, including calibration and background-subtraction
uncertainties (but excluding photon-noise).  In cases where no optical
emission was detected we determined 1-$\sigma$ upper limits.

   There are many locations where UV emission and HI are detected,
but no optical emission is evident (Figure~1): for
example NGC~7771b, the extended SE tidal tail of NGC~5713, or the
plume east of Arp~295b.  In these regions, the H~I column density
is always $\ga$ 2 M$_{\odot}$ pc$^{-2}$.  In a few regions, optical
light is detected with no corresponding UV or HI,  for example at
the end of the SW tail or in the center of the bridge in Arp~295,
or from the red plume extending north from NGC~520.
In these regions, the colors are indicative of old stars, presumably
pulled from the progenitor galaxies during the interaction.

\section{Interpretation} \label{sec:interp}

\subsection{Photometric Ages} \label{sec:ages}

 Luminosity-weighted average ages and extinction for the stellar
populations in UV-bright regions were estimated by comparing
observed FUV$-$NUV and NUV$-$optical colors with reddened
synthetic predictions (Bruzual \& Charlot (2003) models, computed
as described by Bianchi et al. (2004). The observed colors
were corrected for foreground extinction (Schlegel et al. 1998)  
prior to model
fitting. The inclusion of the UV$-$optical color was vital, as it
discounted the possibility that any of the observed FUV$-$NUV
colors were due to hot evolved main-sequence stars with  ages $\ga
10^{9.9}$ yr. In Table~1 we give ages derived under the assumption
of (single) instantaneous burst star formation. These values
represent lower limits to the actual ages (the integrated color of
a mixed-age population is made bluer by the most recent burst,
unless the mass in stars formed is insignificant relative to the
pre-existing total). We emphasize that a good match to the
synthetic colors of a recent burst does not preclude star
formation at earlier times, except in the limit where the latest
burst fully accounts for the bolometric luminosity of a source.
The Arp~295 plume was the only source having colors better fit by
the alternate case of continuous star formation.  In this
scenario, synthetic colors remain blue indefinitely.  It is
plausible that the plume has hosted recurrent bursts (of which
clump 1 and 2 are the most recent) occuring at a rate that makes
star-formation in the region appear psuedo-continuous.  For all
UV-selected regions we also estimated stellar masses (Table 1)
based on the extinction-corrected FUV magnitude and the best
matching synthetic model.

The (FUV$-$NUV, NUV$-$optical) ages for UV-selected regions cover a
substantial range.  NGC~7771 and NGC~5719, in the earliest interactions,
have UV-bright regions ranging from $<$3 Myr (NGC7771b, NGC~5719 clumps) to
$\sim$400Myr. Arp~295 exhibits very recent star formation only in the
eastern plume ($<$ 8Myr), while the rest of the system appears to have
ages between 80Myr and 400Myr. The NGC~520 system (oldest interaction)
shows no evidence for current star formation, and the youngest regions
are  $\sim$300Myr (in UGC~957).  Since the UV-bright regions
identified by GALEX are considerably younger than the dynamical
ages of the tails ($\ga$ 500Myr based on morphology and models,
\citet{mih94,mih96} ), the bursts of star formation must occur
during the evolution of the tidal tails.  {\it It would be
invaluable to have reliable determinations of dynamic histories of
all of these systems, e.g. through targeted models.}

\subsection{Star Formation History in the Tails}\label{sec:evolution}

   The UV observations allow us to  begin tracing
the star formation history of the tidal tails.  In NGC~7771, the
youngest stars are at the end of the H~I tail.  In NGC~5719, the
youngest clump is furthest from the galaxy body, the
diffuse SE tail is considerably younger than the NW tail (which is
younger than the galaxy disk).  In Arp~295, the clumps in the plume
are the youngest, followed by the plume itself; the
long tail that has been ejected from Arp~295a is younger at its SW
extreme and older in the inner regions (the bridge).  In NGC~520,
the youngest stars are in UGC~957.  We also find (Section 2.3), that
the brightest UV emission coincides with HI peaks; examination of
Table~1 shows that these locations are also the bluest (youngest).
Star formation is apparently still occurring
in the possible TDGs, but not in the knots closer to the parent
galaxy.  This
could be because the critical density
takes longer to develop as the distance from the parent
galaxy increases, or because star formation has been
inhibited at the tail / galaxy interface.

   One question of great interest is whether or not the star-forming
clumps in the tidal tails are destined to remain as independent
entities -- that is, as TDGs.  Of the compact UV-bright regions
well-separated from the primary interacting galaxies, several of our
sources have HI mass comparable to confirmed TDGs.  Duc \& Mirabel (1998)
indicate TDGs have HI mass ranging from 10$^{8.3}$ to 10$^{9.8}$
M$_\sun$.  Within our own sample TDG UGC~957 (near NGC~520) has
10$^{8.5}$ M$_\sun$ of HI, and the TDG's in NGC~4038/39 have
10$^{8.2-8.4}$ M$_\sun$ of HI \citep{hib01}.
In comparison, NGC7771b, Arp 295 Clump 1,
and Arp 295 Clump 2 have 10$^{8.7}$, 10$^{8.8}$, and 10$^{8.4}$
M$_\sun$, respectively, of HI coincident with UV emission, making
them plausible TDG candidates.  High-resolution $\lambda21$cm
observations able to gauge the velocity dispersion within these (and
even slightly lower mass condensations like N5719 Clumps 1 and 2,
10$^{7.5-7.8}$ M$_\sun$ of HI) are needed  to determine conclusively
if they are self-gravitating.
Regardless, our early GALEX observations demonstrate the
effectiveness of vacuum UV imaging as a means of identifying TDG
candidates, particularly at the lower limit of the mass function.


\acknowledgments

GALEX is a NASA Small Explorer, launched in April 2003.  We gratefully
acknowledge NASA's support for construction, operation, and science
analysis for the GALEX mission, developed in cooperation with the
Centre National d'Etudes Spatiales of France and the Korean Ministry
of Science and Technology






\clearpage


..
\vskip 3.6truein

{\includegraphics{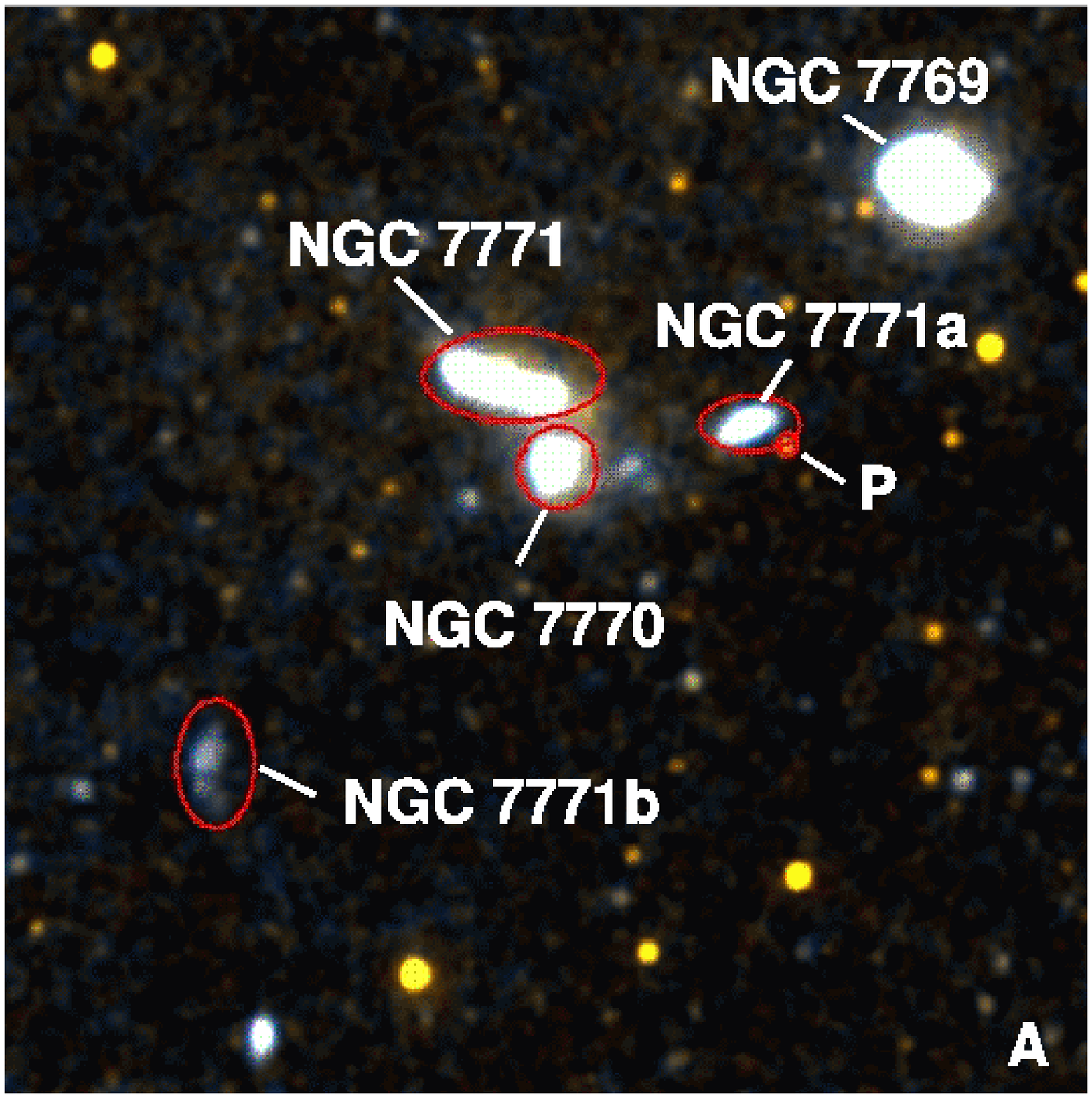}}
{\includegraphics{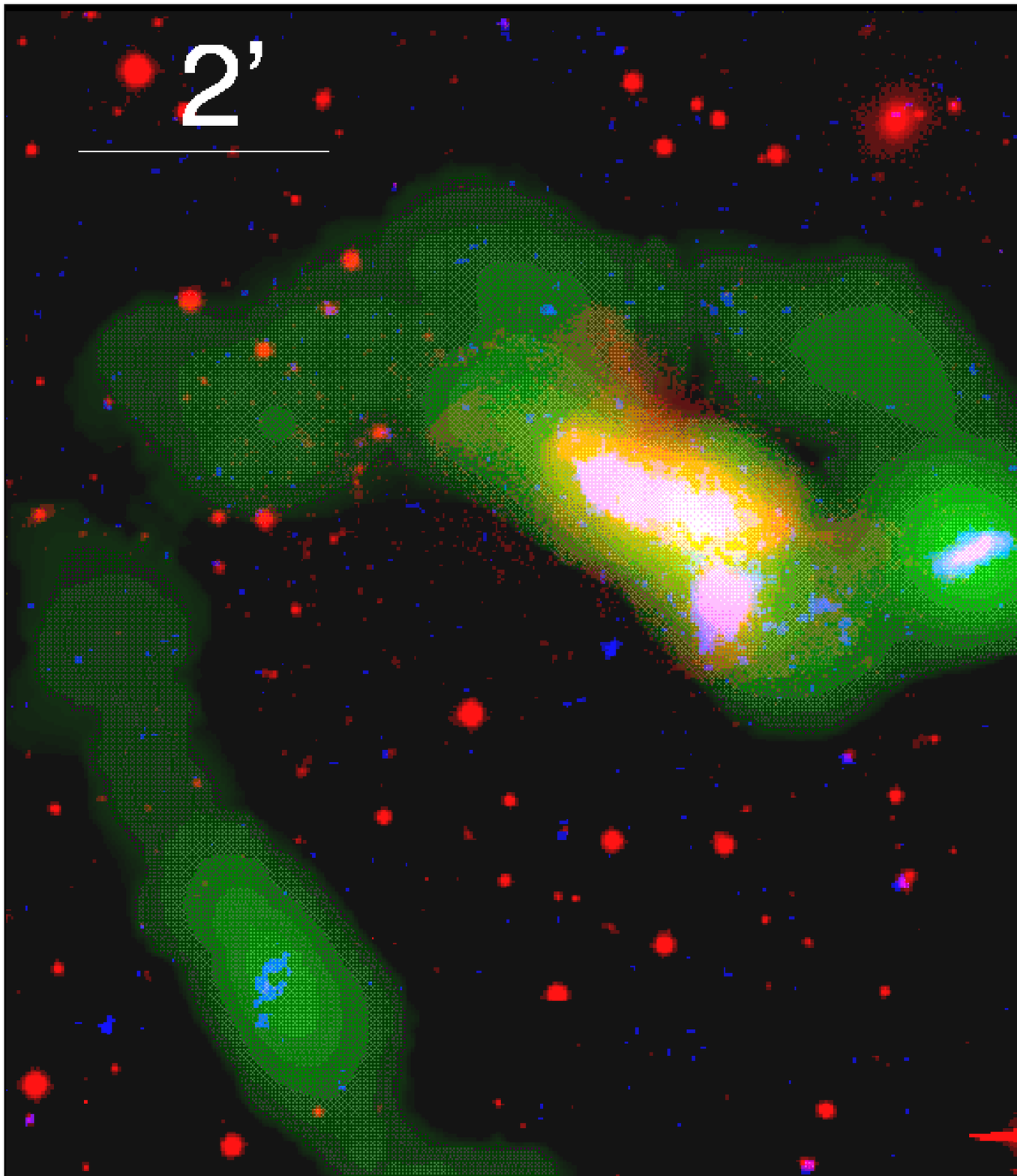}

{\includegraphics{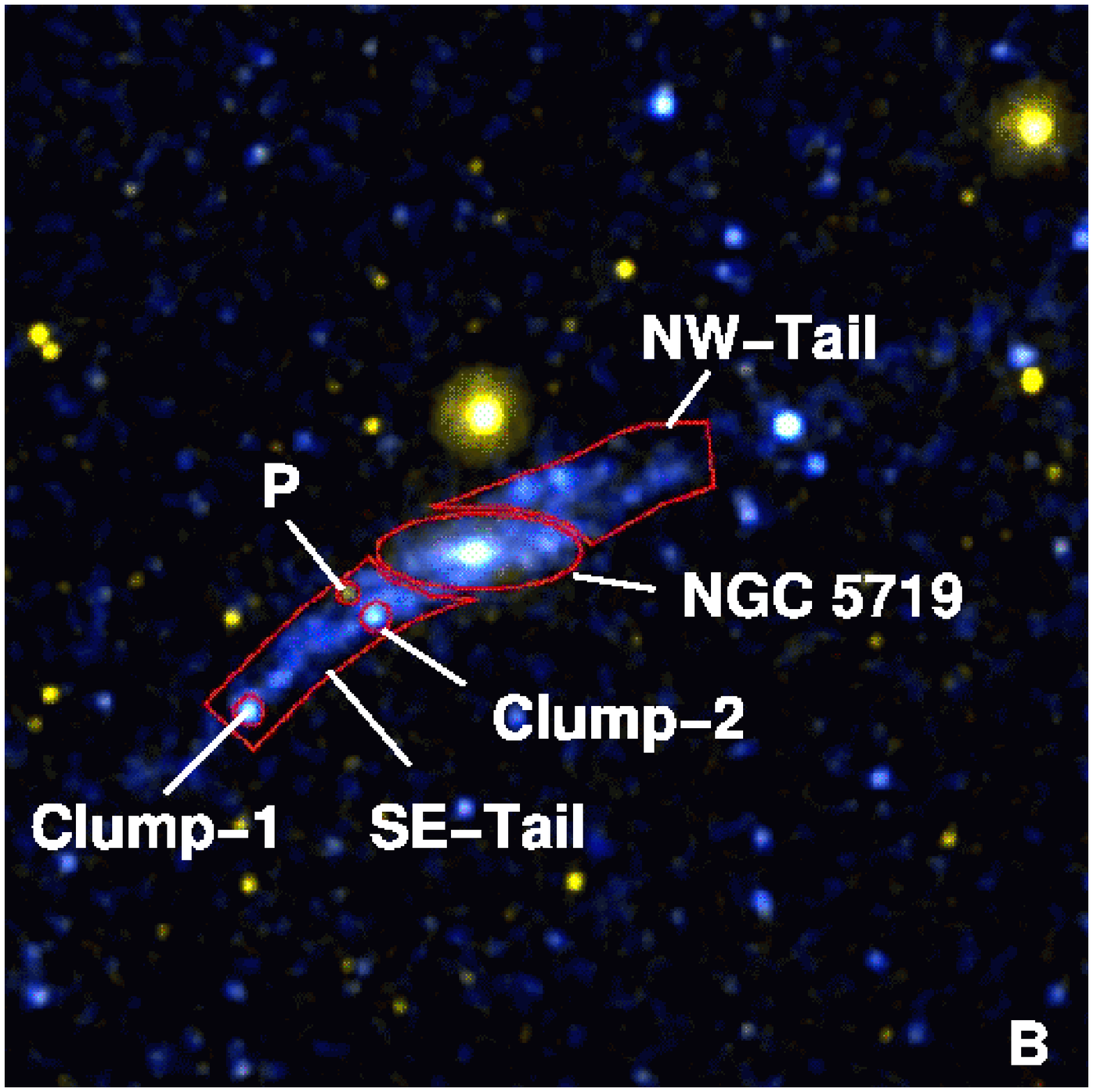}}
\includegraphics{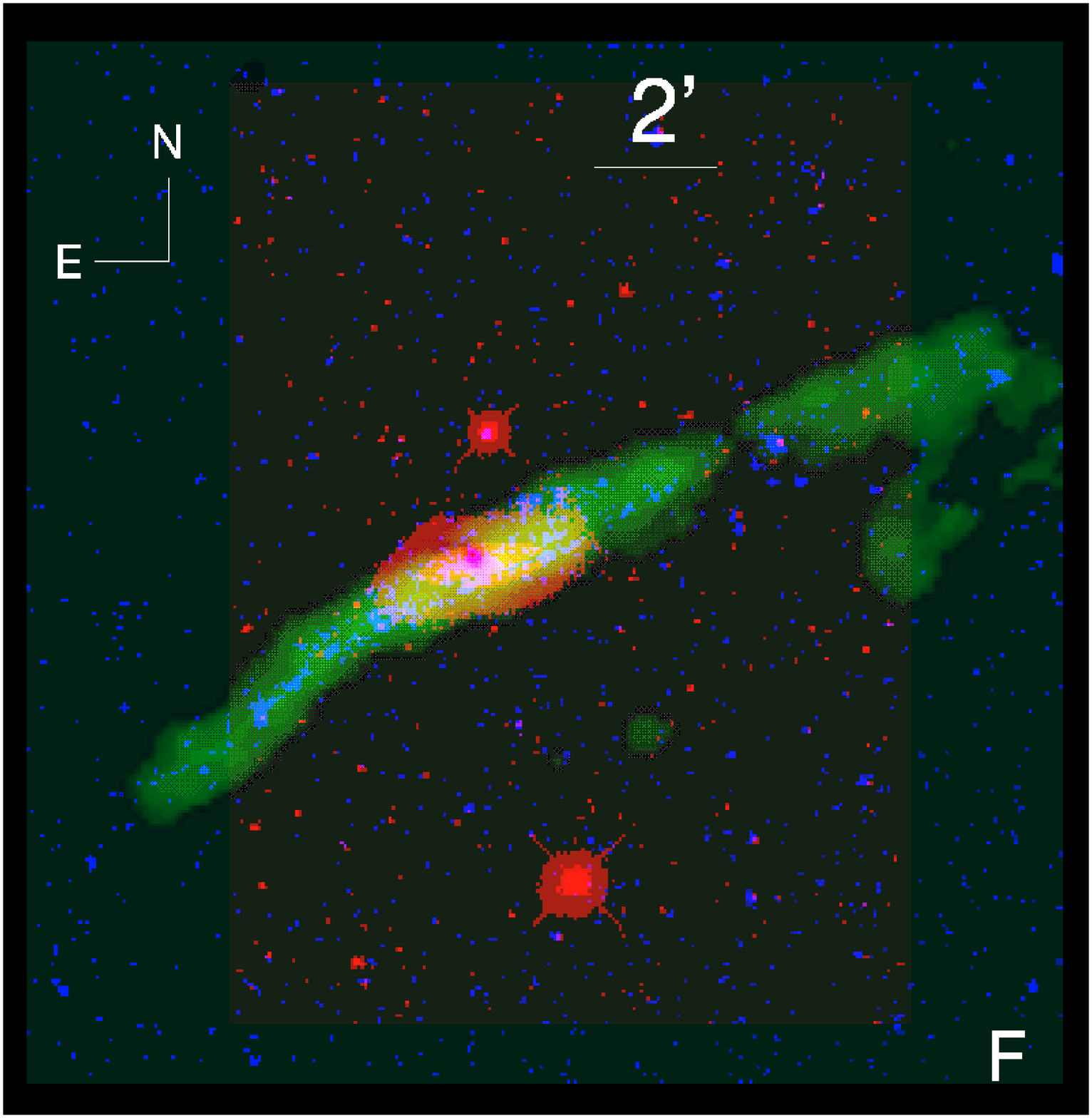}}

{\includegraphics{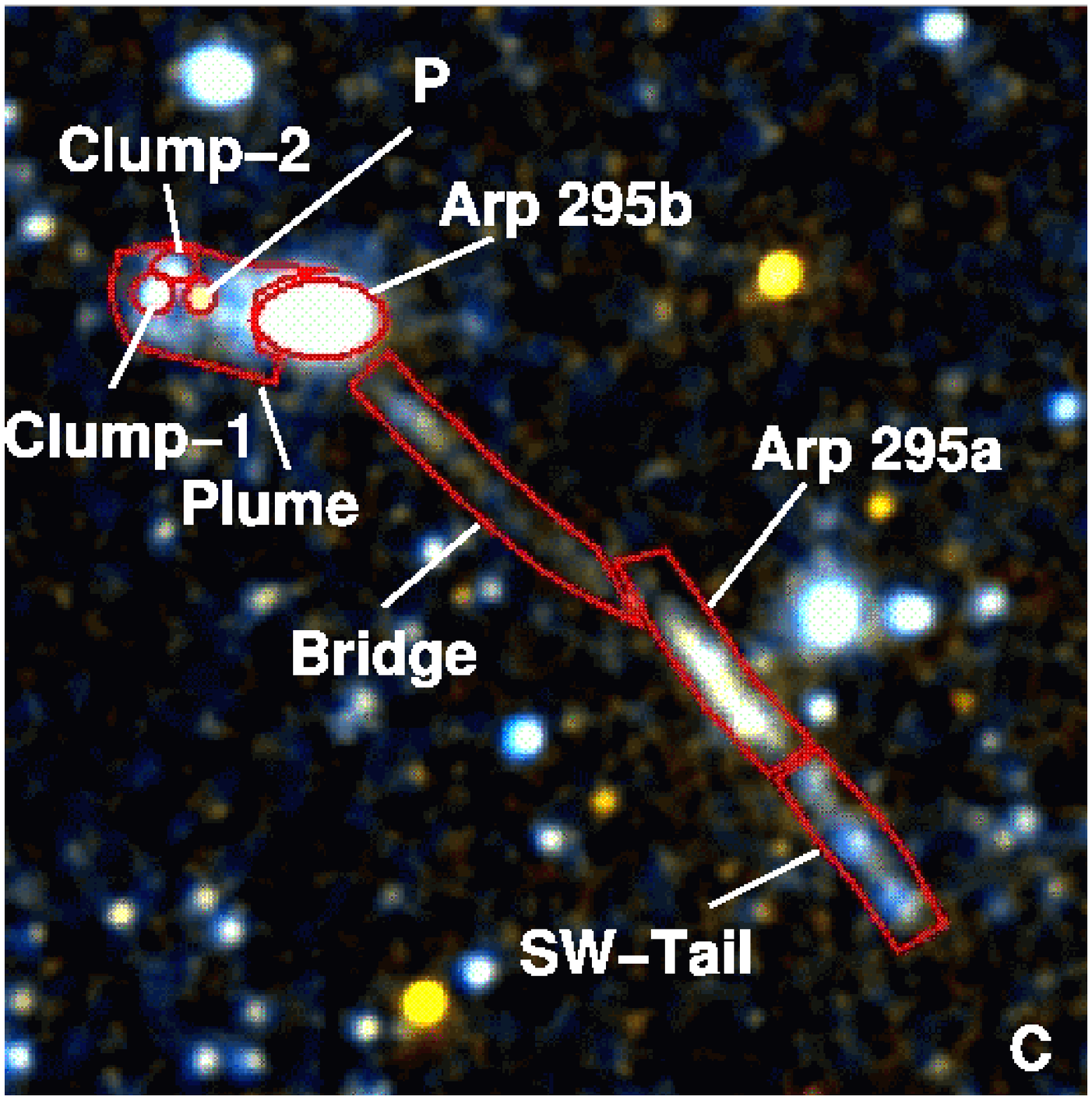}}
{\includegraphics{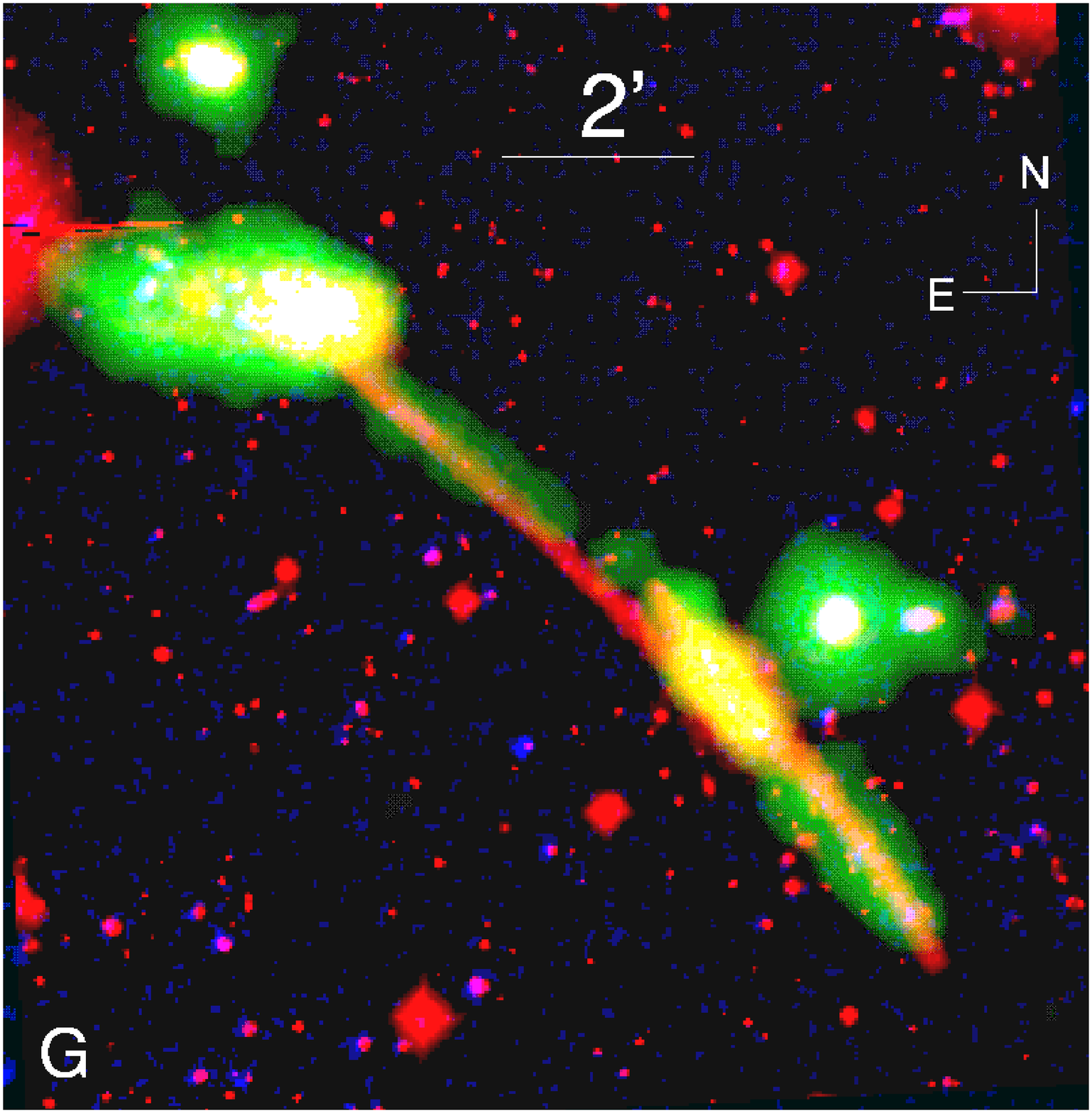}}

{\includegraphics{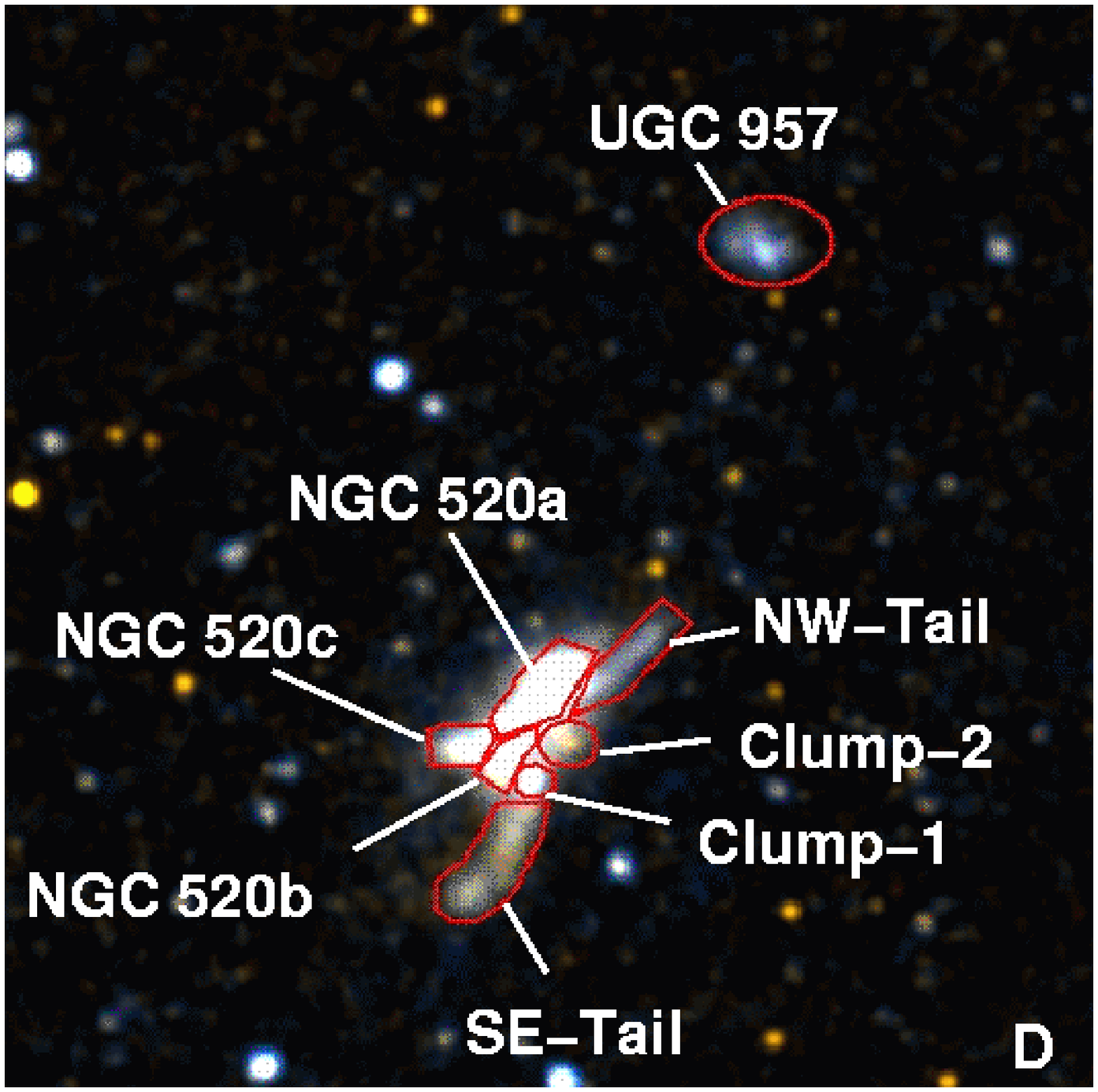}}
{\includegraphics{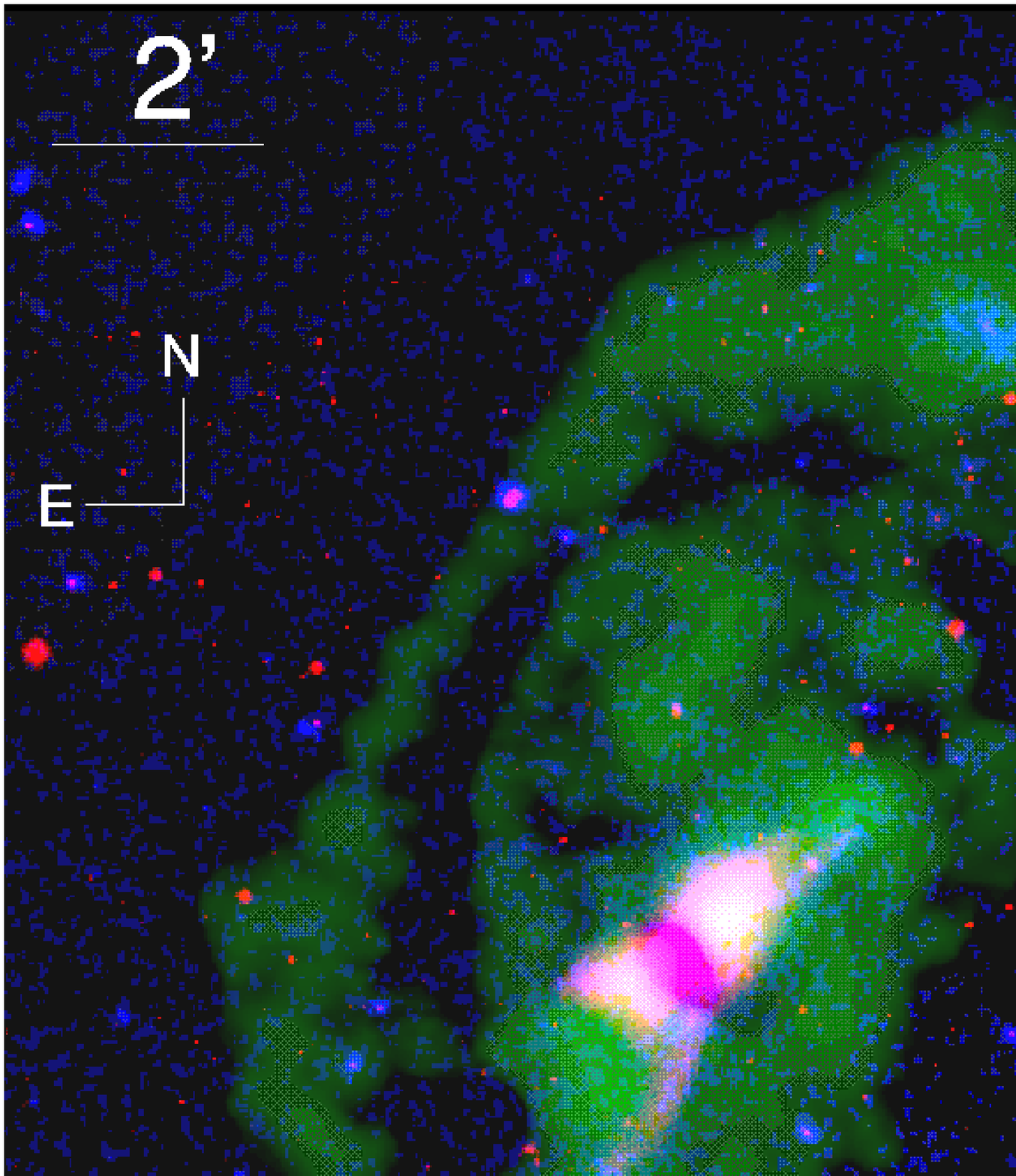}}

{\noindent{\bf Figure 1}
From left to right, panels show tidal tails associated with NGC~7771,
NGC~5719, Arp~295, and NGC~520. Upper panels show FUV (blue), NUV
(orange), and indicate the regions used for photometry. Foreground
stars are labelled ``P'' (``point'') and were excluded from the
photometry. Lower panels show the same galaxy systems in FUV
emission (blue), R or r band (red), and H~I (green).
}

\clearpage

\begin{deluxetable}{llccccccr}
\tabletypesize{\scriptsize}
%
\tablewidth{0pt}
\tablecaption{Observed and Derived Properties of Regions}
\tablehead{
\colhead{System }  &
\colhead{Name } &
\colhead{FUV} &
\colhead{FUV$-$NUV }  &
\colhead{NUV$-r$ \tablenotemark{a} } &
\colhead{Age\tablenotemark{b}} &
\colhead{M$_{\rm stars}$\tablenotemark{c} } &
\colhead{M$_{\rm HI}$ \tablenotemark{d} } &
\colhead{$\Sigma_{\rm HI}$ \tablenotemark{d} } \\
\colhead{} &
\colhead{} & \multicolumn{3}{c}{(before Galactic foreground correction)}  &
\colhead{log (yrs)} &
\colhead{log (M$_{\odot}$)} &
\colhead{log (M$_{\odot}$)} &
\colhead{(M$_{\odot}$pc$^{-2}$)}
}
\startdata
NGC~7769/71&NGC~7770    &16.44$\pm$    0.10&      0.26$\pm$0.13&      
2.86$\pm$0.09&      8.2$^{+0.2}_{-0.2}$&
 9.4$^{+0.1}_{-0.1}$&      8.8 [7.5]  &      2.7 [.1] \\
NGC~7769/71&NGC~7771    &16.90$\pm$    0.05&      0.78$\pm$0.08&      
4.46$\pm$0.06&      8.6$^{+0.1}_{-0.1}$&
10.4$^{+0.1}_{-0.1}$&      9.3 [7.8] &      3.7 [.1] \\
NGC~7769/71&NGC~7771a   &18.38$\pm$    0.07&      0.12$\pm$0.10&      
2.59$\pm$0.07&      8.0$^{+0.1}_{-0.4}$&
 8.4$^{+0.1}_{-0.1}$&      8.8 [7.4]&      3.6 [.1] \\
NGC~7769/71&NGC~7771b   &19.81$\pm$     0.26&     -0.32$\pm$0.40&      
2.43$\pm$0.36&
 
$6.4-7.0$&       7.6$^{+0.1}_{-0.4}$&      8.7 [7.6]&   1.6 [.1]\\
\hline
NGC~5713/19&Disk        &17.96$\pm$    0.08&      0.89$\pm$0.09&      
5.44$\pm$0.04&      8.5$^{+0.1}_{-0.1}$&
10.0$^{+0.1}_{-0.1}$&      9.1 [7.8]&     10.1 [.5]\\
NGC~5713/19&SE~tail     &18.03$\pm$     0.11&     -0.03$\pm$0.14&      
3.27$\pm$0.09&      6.3$^{+0.5}_{-0.2}$&
 8.2$^{+0.1}_{-0.3}$&      9.0 [7.9]&      5.8 [.4]\\
NGC~5713/19&NW~tail     &18.47$\pm$     0.15&      0.25$\pm$0.18&      
3.68$\pm$0.10&      8.0$^{+0.3}_{-1.5}$&
 8.5$^{+0.1}_{-0.2}$&      8.9 [7.9]&      5.0 [.5] \\
NGC~5713/19&Clump~1     &20.11$\pm$     0.19&     -0.25$\pm$0.25&      
1.34$\pm$0.22&      6.4$^{+0.4}_{-0.2}$&
 6.0$^{+0.2}_{-0.4}$&      7.5 [6.6]&      4.5 [.5]\\
NGC~5713/19&Clump~2     &20.07$\pm$     0.18&     -0.24$\pm$0.23&      
2.31$\pm$0.16&
$6.4-6.6$&       6.8$^{+0.1}_{-0.3}$&      7.8 [6.6]&      8.1 [.5] \\
\hline
Arp~295&Arp~295a        &19.74$\pm$     0.19&      1.04$\pm$0.21&      
5.32$\pm$0.09$\dagger$&
8.6$^{+0.1}_{-0.1}$&      10.1$^{+0.1}_{-0.1}$&      9.4 [8.2]&    3.2 [.2]\\
Arp~295&Arp~295b        &16.96$\pm$    0.06&      0.39$\pm$0.09&      
3.03$\pm$0.06$\dagger$&
8.4$^{+0.1}_{-0.1}$&       9.6$^{+0.1}_{-0.1}$&      9.8 [7.9]&    13.1 [.2]\\
Arp~295&SW~tail         &19.98$\pm$     0.21&      0.34$\pm$0.27&      
3.48$\pm$0.18$\dagger$&
8.2$^{+0.3}_{-1.0}$&       8.6$^{+0.1}_{-0.3}$&      9.1 [8.1]&     1.5 [.2]\\
Arp~295&Bridge          &20.87$\pm$     0.58&      1.00$\pm$0.65&      
3.68$\pm$0.28$\dagger$&
8.6$^{+0.1}_{-0.2}$&       8.6$^{+0.1}_{-0.1}$&      8.8 [8.3]&     0.5 [.2]\\
Arp~295&Plume           &18.87$\pm$     0.10&      0.12$\pm$0.15&      
4.53$\pm$0.11$\dagger$&
Cont.&       9.1$^{+0.1}_{-0.1}$&      9.8 [8.3]&      5.8 [.2] \\
Arp~295&Clump~1         &20.56$\pm$     0.15&     -0.09$\pm$0.21&      
1.87$\pm$0.15$\dagger$&
6.5$^{+1.2}_{-0.2}$&       6.8$^{+0.6}_{-0.1}$&      8.8 [7.2] &    6.4 [.2]\\
Arp~295&Clump~2         &21.34$\pm$     0.20&      0.10$\pm$0.23&      
2.36$\pm$0.16$\dagger$&
7.8$^{+0.5}_{-1.3}$&       7.2$^{+0.1}_{-0.2}$&      8.4 [7.1]&    3.4 [.2]  \\
\hline
NGC~520&NGC~520a        &16.83$\pm$    0.05&      0.78$\pm$0.06&      
3.77$\pm$0.14$\dagger$&
8.6$^{+0.1}_{-0.1}$&       9.4$^{+0.1}_{-0.1}$&      8.8 [6.9]&    9.9 [.1] \\
NGC~520&NGC~520b        &19.00$\pm$    0.06&      0.96$\pm$0.06&      
4.24$\pm$0.13$\dagger$&
8.6$^{+0.1}_{-0.1}$&       8.9$^{+0.1}_{-0.1}$&      8.1 [6.5]&    5.0 [.1] \\
NGC~520&NGC~520c        &19.12$\pm$    0.07&      0.71$\pm$0.08&      
3.88$\pm$0.14$\dagger$&
8.5$^{+0.1}_{-0.1}$&       8.5$^{+0.1}_{-0.1}$&      8.3 [6.5]&    7.9 [.1]
\\
NGC~520&UGC~957         &19.42$\pm$    0.09&      0.40$\pm$0.13&      
3.00$\pm$0.16$\dagger$&
8.4$^{+0.1}_{-0.1}$&       7.7$^{+0.1}_{-0.1}$&      8.5 [7.1]&    3.2 [.1] \\
NGC~520&SE~tail         &19.28$\pm$    0.07&      1.16$\pm$0.08&      
3.62$\pm$0.14$\dagger$&
8.6$^{+0.1}_{-0.1}$&       8.4$^{+0.1}_{-0.1}$&      8.4 [7.1]&    3.1 [.2] \\
NGC~520&NW~tail         &19.32$\pm$    0.07&      0.63$\pm$0.08&      
4.01$\pm$0.14$\dagger$&
8.4$^{+0.1}_{-0.1}$&       8.5$^{+0.1}_{-0.1}$&      8.6 [6.9]&    6.2 [.1]\\
NGC~520&Clump~1         &19.91$\pm$    0.09&      0.70$\pm$0.11&      
3.19$\pm$0.14$\dagger$&
8.5$^{+0.1}_{-0.1}$&       7.7$^{+0.1}_{-0.1}$&      7.6 [6.3]&    2.6 [.1]\\
NGC~520&Clump 2         &20.04$\pm$    0.09&      1.09$\pm$0.10&      
4.16$\pm$0.14$\dagger$&
8.6$^{+0.1}_{-0.1}$&       8.4$^{+0.1}_{-0.1}$&      7.9 [6.5]&    3.0 [.1]\\
\enddata
\tablenotetext{a}{NUV$-$R for Arp~295, or NUV$-$V for NGC~520, as indicated by
$\dagger$.}
\tablenotetext{b}{Ages are reported for the case of an instantaneous burst, with
the exception of the Plume in Arp~295 (which was best fit by continuous
star-formation models).  The colors of NGC~7771b and Clump 2 in NGC~5719 were
just outside the grid of
 synthetic predictions (due to observational scatter), so in these cases we
report ages based only on FUV$-$NUV.}
\tablenotetext{c}{The minimum reported 1$\sigma$ error on derived mass is 0.1 
dex, which is larger than the smallest photometric error, but consistent with a 
maximum error of 10\% in assumed distance.}
\tablenotetext{d}{The numbers in brackets are HI detection limits, 
based on a 3$\sigma$ noise/beam in a single channel, over the region 
of interest. All errors are smaller than or equal to 0.1dex, the error 
in assumed distances. }
\label{tab:dertab}
\end{deluxetable}



\begin{thebibliography}{}

\bibitem[Barnes \& Hernquist(1992)]{bar92} Barnes, J. E., \& Hernquist, L. E.,
        1992, \araa, 30, 705
\bibitem[Braine et al.(2001)]{bra01} Braine, Jf., Duc, P.-A., Lisenfeld, U.,
    Charmandaris, v., Vallejo, O., Leon, S.,  \& Brinks, E.,
        2001, \aap, 378, 51
\bibitem[Bruzual \& Charlot(2003)]{bru03} Bruzual, G. \& Charlot, S.,
    2003, \mnras, 344, 1000
\bibitem[Duc \& Mirabel(1998)]{duc98} Duc, P.-A., \& Mirabel, I. F., 1998,
    \aap, 333, 813
\bibitem[Elmegreen \& Efremov(1997)]{elm97} Elmegreen, B. G.
    \& Efremov, Y. N., 1997, \apj, 480, 235
\bibitem[de Grijs et al. (2003)]{deg03} de Grijs, R., Lee, J. T.,
    Clemencia Mora Herrera, M., Fritze-v.Alvensleben, U., \&
    Anders, P., 2003, astroph/0210598.
\bibitem[Haynes \& Giovanelli(1991)]{hay91}  Haynes, M. P., \& Giovanelli, R.,
       1991, \apjs, 77,331
\bibitem[Hibbard \& van Gorkom(1996)]{hib96} Hibbard, J. E. \&
    van Gorkom, J. H. 1996, \aj, 111, 655
\bibitem[Hibbard et al. (2000)]{hib00} Hibbard, J. E., Vacca, W. D., \&
    Yun, M., 2000, \aj 119, 1130,
\bibitem[Hibbard et al.(2001)]{hib01} Hibbard, J., van Gorkon, J.H.,
       Rupen, M. P., Schiminovich, D., 2001,  Gas and Galaxy Evolution,
        ASP conf. series 240, Hibbard, Rupen and van Gorkom eds., 659.
\bibitem[Hibbard et al.(2004)]{hib04} Hibbard, J. E., et al., \apj, ???,
       ??? (this volume).
\bibitem[Knierman et al.(2003)]{kni03} Knierman, K. A., Gallagher, S. C.,
    Charlton, J. C., Hunsberger, S. D., Whitmore, B., Kundu, A.,
    Hibbard, J. E., \& Zaritsky, D.,  2003, \aj, 126, 1227
\bibitem[Kotilainen et al.(2001)]{kot01}  Kotilainen, J. K., Reunanen, J.,
        Laine, S., \& Ryder, S. D., \aa, 366, 439
\bibitem[Langston \& Teuben(2001)]{lan01}  Langston, G., \& Teuben, P.
     Gas and Galaxy Evolution, 2001, ASP conf. series 240, Hibbard, Rupen
     and van Gorkom eds., 862.
\bibitem[Martin et al.(2004)]{mar04} Martin, C. D, et al. 2004, \apj
       (this volume), L1
\bibitem[McMahon et al.(1999)]{McM99} McMahon, R. G. et al., 1999,
        Instrumentation at the ING,
        Walton \& Smartt, New Astronomy Reviews, Elsevier Science
\bibitem[Mihos \& Hernquist(1994)]{mih94} Mihos, J. C. \& Hernquist, L. E.,
        1994, \apj, 427, 112
\bibitem[Mihos \& Hernquist(1996)]{mih96} Mihos, J. C. \& Hernquist, L. E.,
        1996, \apj, 464, 641
\bibitem[Mirabel et al.(1992)]{mir92}  Mirabel, I. F., Dottori, H., \&
    Lutz, D., 1992, \aap, 256, L19
\bibitem[Nordgren et al.(1997)]{nor97} Nordgren, T. E., Chengalur, J. M.,
       Salpeter, E. E., \& Terzian, Y., 1997, \aj, 114, 913
\bibitem[Saviane et al.(2004)]{sav04}  Saviane, I., Hibbard, J. E., \&
   Rich, R. M., 2004,  \aj, 127, 660
\bibitem[Schlegel et al.(1998)]{sch98}  Schlegel, D. J., Finkbeiner, D. P.,
        \& Davis, M, 1998, ApJ, 500, 525.
\bibitem[Schombert et al.(1990)]{sch90}  Schombert, J. M., Wallin, J. F.,
        and Struck-Marcell, C., 1990,  \aj, 99, 497
\bibitem[Stanford \& Balcells(1991)]{sta91} Stanford, S. A., \&
      Balcells, M., 1990, \apj, 355, 59
\bibitem[Toomre \& Toomre(1972]{too72} Toomre,A., \& Toomre, J.,
        \apj, 178, 623
\end{thebibliography}
\end{document}